\documentclass[10pt]{article}
\usepackage[a4paper, left = 2.5cm, right = 2.5cm, top = 2cm, bottom = 2cm]{geometry}
\usepackage{authblk}
\usepackage{amsmath,amssymb,amsfonts}
\usepackage{bm}
\usepackage[dvipsnames]{xcolor}
\usepackage{tabularx}
\usepackage[colorlinks=true,urlcolor=black,linkcolor=black, citecolor=black]{hyperref}
\usepackage{caption}
\usepackage{subcaption}
\usepackage{setspace}
\usepackage{afterpage}
\usepackage{sectsty}

\DeclareMathOperator*{\argmin}{argmin}

\usepackage[final, todonotes={textsize=small},defaultcolor=BrickRed]{changes}
\definechangesauthor[name=GS, color=blue]{GS}
\definechangesauthor[name=FB, color=red]{FB}
\definechangesauthor[name=JN, color=green]{JN}
\definechangesauthor[name=ME, color=orange]{ME}

\usepackage[backend=bibtex,style=ieee,doi=true,url=true]{biblatex}
\title{\sffamily\textbf{Joint estimation of activity, attenuation and motion in respiratory-self-gated time-of-flight PET}}

\author[1]{Masoud Elhamiasl}
\author[1]{Frederic Jolivet}
\author[1]{Ahmadreza Rezaei}
\author[2]{Michael Fieseler}
\author[2]{Klaus Sch\"afers}
\author[1]{Johan Nuyts}
\author[1]{Georg Schramm\thanks{\added{corresponding author email: georg.schramm@kuleuven.be}}}
\author[3]{Fernando Boada}
\affil[1]{Department of Imaging and Pathology, KU Leuven, Belgium}
\affil[2]{European Institute for Molecular Imaging (EIMI), Universit\"at M\"unster, Germany}
\affil[3]{Department of Radiology, Stanford School of Medicine, US}

\date{\today}

\begin{document}
\allsectionsfont{\sffamily}
\maketitle

\begin{abstract}
\textbf{\replaced{Objective}{Motivation}:} 
Whole-body Positron Emission Tomography (PET) imaging is often hindered by respiratory 
motion during acquisition, causing significant degradation in the quality of 
reconstructed activity images. 
An additional challenge in PET/CT imaging arises from the respiratory phase mismatch 
between CT-based attenuation correction and PET acquisition, 
leading to attenuation artifacts. 
To address these issues, we propose two new, purely data-driven methods 
for the joint estimation of activity, attenuation, and motion in respiratory 
self-gated time-of-flight (TOF) PET. 
These methods enable the reconstruction of a single activity image 
free from motion and attenuation artifacts.

\textbf{\replaced{Approach}{Methods}:} 
The proposed methods were evaluated using data from the anthropomorphic Wilhelm 
phantom acquired on a Siemens mCT PET/CT system, as well as three clinical 
[\textsuperscript{18}F]FDG PET/CT datasets acquired on a GE DMI PET/CT system. 
Image quality was assessed visually to identify motion and attenuation artifacts.
Lesion uptake values were quantitatively compared across reconstructions 
without motion modeling, with motion modeling but ``static'' attenuation correction, 
and with our proposed methods.

\textbf{\replaced{Main results}{Results}:} 
For the Wilhelm phantom, the proposed methods delivered image quality closely matching 
the reference reconstruction from a static acquisition. The lesion-to-background 
contrast for a liver dome lesion improved from 2.0 (no motion correction) to 5.2 
(using our proposed methods), matching the contrast from the static acquisition (5.2). 
In contrast, motion modeling with ``static'' attenuation correction yielded a lower 
contrast of 3.5. 
In patient datasets, the proposed methods successfully 
reduced motion artifacts in lung and liver lesions and mitigated attenuation 
artifacts, demonstrating superior lesion to background separation.

\textbf{\replaced{Significance}{Conclusion}:} 
Our proposed methods enable the reconstruction of a single, high-quality activity 
image that is motion-corrected and free from attenuation artifacts, without the 
need for external hardware.

\end{abstract}

\section{Introduction}
\label{sec:intro}
Positron Emission Tomography (PET) is an imaging modality used to visualize and 
quantify metabolic processes within the body. 
The duration of a whole-body PET scan typically spans several minutes. During 
such acquisitions, respiratory motion of the patient is inevitable, leading to 
substantial degradation in the quality of the reconstructed activity 
image~\cite{lamare2022pet}. Current approaches for mitigating motion artifacts 
involve the reconstruction and alignment of respiratory-gated list-mode data.

Despite its effectiveness, respiratory gating is not commonly employed in 
clinical practice, largely due to the use of external hardware (e.g., pressure 
belts) for acquiring a respiratory gating signal, which complicates an efficient 
clinical workflow. 
Nevertheless, recent studies have demonstrated that data-driven extraction of 
respiratory gating signals from the acquired raw data is 
feasible~\cite{thielemans2011device}, potentially facilitating the routine 
application of respiratory-gated reconstructions in clinical settings.

A practical challenge in whole-body PET/CT imaging arises from the use of 
a single static CT image for attenuation correction, 
which is typically acquired during  breath-hold and thus not phase-matched 
to most of the respiratory gates of the PET acquisition. 
This mismatch results in well-documented attenuation artifacts 
in the reconstructed images, such as the so-called ``banana artifact'' 
observed in the liver dome.

Given that joint estimation of activity and attenuation using time-of-flight 
(TOF) data has been demonstrated~\cite{defrise2012time, rezaei2012mlaa}, 
it is appealing to pursue a joint estimation of activity and 
``respiratory-gate-matched'' attenuation. 
However, performing this estimation on a gate-by-gate basis presents 
significant challenges due to the limited number of counts available 
in each respiratory gate.

In this work, focusing on the reconstruction of whole-body TOF PET data 
affected by respiratory motion, our aim is to jointly estimate 
``gate-matched'' attenuation, respiratory motion vector fields, and a unified 
activity image that is free from motion and attenuation artifacts, without 
relying on external hardware for motion signal extraction. Specifically, 
our approach involves:
\begin{enumerate}
    \item purely data-driven extraction of the respiratory gating signal 
          and definition of respiratory gates directly from the acquired 
          list-mode data,
    \item estimation of respiratory phase-matched attenuation for each gate,
    \item estimation of motion vector fields between a reference gate and 
          all other gates,
    \item simultaneous reconstruction of a single 3D activity image from 
          the entire PET emission data, using a forward model that 
          incorporates the motion vector fields and the estimated 
          ``gate-matched'' attenuation sinograms.
\end{enumerate}
To achieve this aim, we propose two methods. Our first approach, a simple 
hybrid method, relies on estimating the ``gate-matched'' attenuation sinograms 
and motion vector fields from initial gate-by-gate MLACF 
\added{(Maximum Likelihood Attenuation Correction Factors)}
reconstructions~\cite{rezaei2014mlacf}, followed by an OS-MLEM 
\added{(Maximum Likelihood Expectation Maximization with Ordered Subsets)}
reconstruction 
of a single activity image from the complete dataset using the estimated 
attenuation sinograms and motion vector fields.

The second, more complex method aims to jointly estimate attenuation, 
motion vector fields, and a single activity image by adapting the 
Alternating Direction Method of Multipliers 
(ADMM)~\cite{boyd2011distributed}.

We evaluate the performance of our proposed methods against more conventional 
approaches, using acquisitions of the anthropomorphic Wilhelm phantom 
\cite{bolwin2018anthropomorphic}, as well as three clinical whole-body 
[\textsuperscript{18}F]FDG acquisitions performed on a Siemens Biograph mCT 
and a GE DMI PET/CT, respectively.

The remainder of this article is structured as follows: 
Section \ref{sec:theory} presents the formal problem statement and theory, 
along with an overview of related work. 
Section \ref{sec:mat} describes the experiments conducted to evaluate our 
proposed methods, with the corresponding results detailed in 
Section \ref{sec:res}. Finally, a discussion of all methods, results,
and limitations is provided in Section \ref{sec:dis}.

\section{Problem statement, theory and related work}
\label{sec:theory}
In this work, our objective is to reconstruct a single activity image 
$\bm{\lambda}$ from a series of data-driven respiratory-gated emission 
sinograms $\{\bm{y}^k\}$. Additionally, we assume the availability of a 
``static'' attenuation image $\bm{\mu}$ and the corresponding derived 
attenuation sinogram $\tilde{\bm{a}}$, which may be based, for example, on a 
CT or MR scan acquired during breath-hold. It should be noted, however, that 
this static attenuation image is not phase-matched to any of the 
respiratory gates $k$, resulting in the respiratory phase-matched attenuation 
sinograms $\{\bm{a}^k\}$ being unknown.

Using a series of unknown non-rigid image warping operators 
$\{\bm{S}^k\}$, which transform an image from a reference gate to another 
gate $k$, and unknown gate-specific attenuation sinograms $\{\bm{a}^k\}$, 
we can express the forward model for the expectation of the TOF emission 
data in gate $k$ as
\begin{equation}
    \bar{y}^k_{it}(\bm{\lambda}) = a_i^k \left(\bm{P} (\bm{S}^k{\bm{\lambda}}) \right)_{it} 
    + r^k_{it} \label{eq:fwd_model} \ ,
\end{equation}
where $i$ represents a geometrical line of response (LOR) connecting two 
detectors, $t$ denotes a time-of-flight (TOF) bin along $i$, $\bm{P}$ 
is a TOF forward projector that includes all corrections, such as 
normalization, but excludes attenuation factors, and
$r_{it}^k$ represents the expected number of scattered and random coincidences.

In cases where the image warping operators $\{\bm{S}^k\}$ and the 
gate-specific attenuation sinograms $\{\bm{a}^k\}$ are known a priori, a 
motion-corrected image reconstruction can be obtained by solving the 
following optimization problem:
\begin{equation}
    \hat{\bm{\lambda}} \in \argmin_{\bm{\lambda}} \sum_k
    \underbrace{\sum_i \sum_t \bar{y}^k_{it}(\bm{\lambda}) - y^k_{it}\log(\bar{y}^k_{it}(\bm{\lambda})) }_
    {\mathcal{D}_k(\bm{a}^k, \bm{S}^k\bm{\lambda})} 
    + \mathcal{R}_1(\bm{\lambda}) \label{eq:problem} \ ,
\end{equation}
where $\mathcal{D}_k$ represents the negative Poisson log-likelihood for 
gate $k$, and $\mathcal{R}_1$ is a regularizer acting on the activity 
image $\bm{\lambda}$.

Chun and Fessler~\cite{chun2013} have demonstrated that motion-corrected 
image reconstruction outperforms post-reconstruction image alignment 
approaches~\cite{klein1997, dawood2006, bai2009}, provided that the exact 
image warping operators $\{\bm{S}^k\}$ and all attenuation sinograms 
$\{\bm{a}^k\}$ are known. However, as mentioned previously, this is often 
not the case in standard clinical whole-body PET imaging, where motion 
tracking, modeling, or gated attenuation images are typically unavailable.

Since it is well established that TOF PET data contain information about 
attenuation~\cite{defrise2012time} and that the registration of 
gate-by-gate reconstructions can provide insights into the image warping 
operators, it is compelling to attempt a joint estimation of the image 
warping operators $\{\bm{S}^k\}$ and all attenuation sinograms 
$\{\bm{a}^k\}$ in addition to the single activity image $\bm{\lambda}$ 
based on the gated acquired data $\{\bm{y}^k\}$. We refer to this approach as 
joint image reconstruction including estimation of motion and attenuation 
(\textbf{JRMA}), which can be formulated as the following optimization 
problem:
\begin{equation}
    \hat{\bm{\lambda}}, \{\hat{\bm{S}}_k\}, \{\hat{\bm{a}}_k\} \in 
    \argmin_{\bm{\lambda},\{\bm{S}^k\},\{\bm{a}^k\}}
    \sum_k \underbrace{\sum_i \sum_t \bar{y}^k_{it}(\bm{\lambda}) - 
    y^k_{it}\log(\bar{y}^k_{it}(\bm{\lambda})) }_
    {\mathcal{D}_k(\bm{\bm{a}^k, \bm{S}^k \lambda})} 
    + \mathcal{R}_1(\bm{\lambda}) + \mathcal{R}_2(\bm{a}^k) +
    \mathcal{R}_3(\bm{S}^k) \ . \label{eq:joint_problem}
\end{equation}
Solving \eqref{eq:joint_problem} is inherently challenging due to the 
following reasons:
\begin{enumerate}
    \item The joint optimization problem is not jointly convex in 
          $\hat{\bm{\lambda}}, \{\hat{\bm{S}}_k\}, \{\hat{\bm{a}}_k\}$.

    \item The count statistics in each respiratory gate are typically 
          very low, complicating the estimation of attenuation sinograms 
          $\{\bm{a}^k\}$ based on TOF data and hindering the 
          data-driven motion estimation process.
\end{enumerate}

%%%%
In 2018, Lu et al.~\cite{lu2018} demonstrated that gate-by-gate PET 
reconstructions using MLACF~\cite{rezaei2014mlacf} yielded the most accurate 
estimation of motion vector fields between gates compared to 
reconstructions that either ignored attenuation or utilized static 
CT-based attenuation. This finding motivated us to incorporate 
MLACF-based estimation of gated attenuation sinograms in the joint 
estimation problem, with the expectation that this would lead to more 
accurate motion estimates as well as reconstructions free from respiratory 
attenuation artifacts.

In 2016, Bousse et al.~\cite{bousse2016} presented an algorithm for joint 
reconstruction and motion estimation that included alignment of a single 
attenuation image, which was not phase-matched to any of the respiratory 
gates (JRM). This algorithm was evaluated using both simulated non-TOF data 
and real non-TOF patient data acquired on a GE Discovery STE and a Siemens 
Biograph mMR~\cite{bousse2017}. In their discussion, the authors noted that:
\begin{itemize}
    \item ``... JRM with misaligned $\mu$-map requires a large number of 
          iterations, indicating that for non-TOF-PET data the problem is 
          ill-posed.``
          
    \item ``However, since JRM uses a known $\mu$-map, it does not suffer 
          from the same cross-talk issues as joint reconstruction of 
          activity and attenuation in non-TOF-PET ...''
          
    \item ``TOF-PET data would likely accelerate convergence of JRM, but we 
          leave this for future work.''
\end{itemize}
In a subsequent fast track communication~\cite{bousse2016b}, Bousse et al. 
indeed demonstrated that, based on simulated data, TOF significantly 
accelerates the convergence of JMRA. However, the proposed algorithm was 
not evaluated using real clinical TOF data.

\paragraph{Main novelties of this work:} Building on the work 
of Bousse et al., as well as our preliminary results published in the 
proceedings~\cite{jolivet2022}, we present two novel algorithms for JMRA, 
which include direct MLACF-based estimation of gated attenuation sinograms 
from the acquired TOF emission data. 

Furthermore, by factorizing the gated attenuation sinograms into the product 
of a static CT-based attenuation sinogram and gate-specific correction 
factors, and by employing an intensity prior that favors correction factors 
close to unity, we also address the inevitable scaling issue inherent in 
joint estimation of activity and attenuation.
\added{With MLACF, we avoid the need for reconstructing attenuation images, making the approach more computationally efficient. 
However, this method does not allow constraining the attenuation sinograms to correspond to projections of a single attenuation
image deformed with the same deformations as the activity images.}

We evaluate our proposed algorithms using acquisitions of the anthropomorphic 
Wilhelm phantom, as well as real patient data acquired on two state-of-the-art 
TOF PET/CT systems.

%%%%%%%
\subsection{MLEM for JR}

In the hypothetical scenario where all motion deformation operators 
$\{\bm{S}^k\}$ and gated attenuation sinograms $\{\bm{a}^k\}$ are known 
a priori, and in the absence of the image regularizer 
$\mathcal{R}_1(\bm{\lambda})$, the optimization problem \eqref{eq:problem} 
can be solved using the Maximum Likelihood Expectation Maximization (MLEM) 
algorithm, using the iterative update
\begin{equation}
    \bm{\lambda}^{(n+1)} = \frac{\bm{\lambda}}{\sum_k \left(\bm{S}^k\right)^T 
    \bm{P}^T \bm{a}^k} \sum_k \left(\bm{S}^k\right)^T
    \bm{P}^T \bm{a}^k \frac{\bm{y}^k}{\bar{\bm{y}}^k(\bm{\lambda^{(n)}})} \ ,
    \label{eq:jr_mlem}
\end{equation}

where the forward model $\bar{\bm{y}}^k(\bm{\lambda^{(n)}})$ is given in \eqref{eq:fwd_model}.

%%%%%%%%%%%
\subsection{Hybrid JMRA using Gate-by-Gate MLACF-Based Estimation of Motion and Attenuation}

The first proposed method, termed ``hybrid JMRA'', involves the following steps:
\begin{enumerate}
    \item Perform a gate-by-gate MLACF reconstruction~\cite{rezaei2014mlacf} 
          to obtain 
          \added{gate-dependent multiplicative correction factors to
          the static attenuation sinogram $\bm{\tilde{a}}$ using the updates
          defined in appendix \ref{sec:app_mlacf}} resulting in  
          phase-matched attenuation sinograms 
          $\{\bm{a}^k_\text{MLACF}\}$, as well as intermediate 
          gate-by-gate reconstructions $\{\bm{\lambda}^k_\text{MLACF}\}$. 

    \item Non-rigidly align all $\{\bm{\lambda}^k_\text{MLACF}\}$ to the 
          MLACF reconstruction of the reference gate in order to 
          estimate all motion warping operators $\{\bm{S}^k\}$.

    \item Perform joint reconstruction (JR) with fixed 
          $\{\bm{a}^k\}$ and $\{\bm{S}^k\}$ estimated in two previous steps 
          using \eqref{eq:jr_mlem}
          to obtain a single activity image $\bm{\lambda}$ from the 
          respiratory-gated emission sinograms.
\end{enumerate}

%%%%%%%%%%%%%%%%%%

\subsection{ADMM-Based Reconstruction for JR}

Before addressing the more complex problem of true JRMA in our second
proposed method, we first consider 
a simplified reconstruction problem where all motion deformation operators 
$\{\bm{S}^k\}$ and gated attenuation sinograms $\{\bm{a}^k\}$ are known 
a priori, and where a regularizer $\mathcal{R}_1(\bm{\lambda})$ is present:
\begin{equation}
    \hat{\bm{\lambda}} \in
    \argmin_{\bm{\lambda}} \sum_k \mathcal{D}_k(\bm{a}^k, \bm{S}^k\bm{\lambda}) 
    + \mathcal{R}_1(\bm{\lambda}) \label{eq:jr_problem} \ ,
\end{equation}
which, due to the presence of the regularizer, cannot be solved in an 
``uncoupled'' (gate-by-gate) manner. However, by introducing the additional 
constraint variables $\bm{z}^k = \bm{S}^k \bm{\lambda}$, we can reformulate 
the problem as
\begin{equation}
    \hat{\bm{\lambda}} \in \argmin_{\substack{\bm{z}^k, \bm{\lambda}\\ \text{s.t}~\bm{z}^k = \bm{S}^k\bm{\lambda}}}
    \sum_k \mathcal{D}_k(\bm{a}^k, \bm{z}^k) + \mathcal{R}_1(\bm{\lambda})
\end{equation}
and optimize the corresponding \deleted{(scaled)} augmented Lagrangian \cite{boyd2011distributed}
\added{
\begin{equation}
    \mathcal{L}_\rho (\bm{\lambda}, \{\bm{z}^k\}, \{\bm{w}^k\}) = \sum_k
    \Big(\mathcal{D}_k(\bm{z}^k, \bm{a}^k) + 
    (\bm{w}^k)^T \left( \bm{z}^k - \bm{S}^k \bm{\lambda} \right) + 
    \frac{\rho}{2} \Vert \bm{z}^k - \bm{S}^k \bm{\lambda} \Vert_2^2 \Big) 
    + \mathcal{R}_1(\bm{\lambda}) \ ,
\end{equation}
which can be re-written into}
\begin{equation}
    \mathcal{L}_\rho (\bm{\lambda}, \{\bm{z}^k\}, \{\bm{u}^k\}) = \sum_k
    \underbrace{\Big(\mathcal{D}_k(\bm{z}^k, \bm{a}^k) + \frac{\rho}{2} 
    \Vert \bm{z}^k - \bm{S}^k \bm{\lambda} + \bm{u}^k \Vert_2^2 
    - \frac{\rho}{2} \Vert \bm{u}^k \Vert_2^2 \Big)}_{L^k_\rho(\bm{\lambda}, \bm{z}^k, \bm{u}^k)} + \mathcal{R}_1(\bm{\lambda}) \ ,
\end{equation}
\added{by using the scaled dual variable $\bm{u}^k = \bm{w}^k / \rho$.}

The complex problem in \eqref{eq:jr_problem} can then be solved iteratively using 
the Alternating Direction Method of Multipliers (ADMM)~\cite{boyd2011distributed}
which splits the problem into the following three ``simpler'' subproblems
\begin{align}
    \bm{z}^{k,(n+1)}     & = \argmin_{\bm{z}}~L^k_\rho \Big(\bm{\lambda}^{(n)}, \bm{z}, \bm{u}^{k,(n)}\Big) 
    \hspace{1cm} \forall \text{ gates } k \label{eq:admm_1} \\
    \bm{\lambda}^{(n+1)} & = \argmin_{\bm{\lambda}}~\mathcal{L}_\rho \Big(\bm{\lambda}, \{\bm{z}^{k,(n+1)}\}, \{\bm{u}^{k,(n)}\}\Big)  
    \label{eq:admm_2} \\
    \bm{u}^{k,(n+1)}     & = \bm{u}^{k,(n)} + \bm{z}^{k,(n+1)} - \bm{S}^k \bm{\lambda}^{(n+1)} 
    \hspace{1cm} \forall \text{ gates } k
\end{align}
Note that subproblem \eqref{eq:admm_1} can be solved on a gate-by-gate basis 
and involves optimizing the Poisson data fidelity term along with a quadratic 
penalty term, which can be efficiently handled using methods such as 
De Pierro's algorithm~\cite{de1995modified}. The subsequent subproblem 
\eqref{eq:admm_2} is related to the classical denoising problem that can be solved 
effectively using various solvers from convex optimization theory.

%%%%%%%%%%%%%%%%%%
\subsection{JMRA Based on a Modified ADMM Algorithm}

To jointly estimate the unknown motion deformation operators $\{\bm{S}^k\}$, 
as well as the gate-by-gate attenuation sinograms $\{\bm{a}^k\}$, we treat 
them as variables in the augmented Lagrangian:

\begin{equation}
    \begin{split}
        \mathcal{L}_\rho (\bm{\lambda},\{\bm{z}^k\},\{\bm{u}^k\},\{\bm{a}^k\},\{\bm{S}^k\}) = \sum_k \Big(\mathcal{D}_k(\bm{z}^k, \bm{a}^k) +
        \frac{\rho}{2} \Vert \bm{z}^k - \bm{S}^k \bm{\lambda} + \bm{u}^k \Vert_2^2 - \frac{\rho}{2} \Vert \bm{u}^k \Vert_2^2 \Big) \\
        + \mathcal{R}_1(\bm{\lambda}) + \mathcal{R}_2(\bm{a}^k) + \mathcal{R}_3(\bm{S}^k) \ .
    \end{split}
\end{equation}

To optimize this more complex and non-jointly convex augmented Lagrangian, 
we propose a heuristic modification of the original ADMM algorithm, dividing 
the problem into four ``simpler'' subproblems in each iteration:

\begin{align}
    & \text{subproblem 1: } \bm{z}^{k,(n+1)} = \argmin_{\bm{z}^k}~\mathcal{L}_\rho 
    \Big(\bm{\lambda}^{(n)}, \bm{z}^k, \bm{u}^{k,(n)}, \bm{a}^{k,(n)}, \bm{S}^{k,(n)} \Big) 
    \quad \forall \text{ gates } k \label{eq:sp_1} \\
    & \text{subproblem 2: } \bm{a}^{k,(n+1)} = \text{MLACF attenuation update based on } 
    \bm{S}^{k,(n)}, \bm{\lambda}, \ \bm{a}^{k,(n)} \label{eq:sp_2} \\
    & \text{subproblem 3: } \bm{\lambda}^{(n+1)} = \argmin_{\bm{\lambda}}~\mathcal{L}_\rho 
    \Big(\bm{\lambda}, \bm{z}^{k,(n+1)}, \bm{u}^{k,(n)}, \bm{a}^{k,(n+1)}, \bm{S}^{k,(n)} \Big) 
    \quad \forall \text{ gates } k \label{eq:sp_3} \\
	 & \text{subproblem 4: } \bm{S}^{k,(n+1)} = \text{update based on non-rigid registration of } \bm{z}^{k,(n+1)} \nonumber \label{eq:sp_4}              \\
	 & \hspace{5cm} \text{ to $\bm{z}^{k,(n+1)}$ of the reference gate} \\
    & \text{subproblem 5: } \bm{u}^{k,(n+1)} = \bm{u}^{k,(n)} + \bm{z}^{k,(n+1)} - 
    \bm{S}^{k,(n+1)} \bm{\lambda}^{(n+1)} \quad \forall \text{ gates } k
\end{align}

Our approaches to solve the first four subproblems in every outer iteration of the 
modified ADMM algorithm for JRMA are detailed in the appendix \ref{app:one}.
\added{It is important to reiterate that the MLACF update in subproblem 2 estimates multiplicative gate-by-gate correction sinograms, 
which are applied on top of the existing static attenuation sinogram to model
the gate dependent attenuation - see appendix \ref{sec:app_mlacf} for details.}
Note that, to simplify the tuning of the Lagrangian penalty parameter $\rho$, 
we consistently use rescaled forward operators such that the norm of the 
operator is normalized to unity in all reconstructions performed in this work.

In subproblem 5, an important consideration arises: whether to align 
$\bm{z}^k$ to $\bm{\lambda}$ or instead align $\bm{z}^k$ to the 
$\bm{z}$ image of the reference gate. 
Since $\bm{\lambda}$ may not have converged to the motion-free image in the 
reference position during the early iterations, 
we opt to align to the $\bm{z}$ image of the reference gate instead. 
This approach ensures  greater stability in the iterative process.

\section{Materials and Methods}
\label{sec:mat}
\subsection{Data-driven estimation of the respiratory gating signal and gate definition}\label{sec:resp_signal}

For any reconstruction algorithm including \added{periodic} motion modeling and compensation,
an accurate estimate of the respiratory motion (gating) signal is required. 
This 1D time signal is used to define the respiratory gates. 

Since a respiratory signal based on external hardware motion trackers is typically unavailable in most clinical PET acquisitions, and as we are aiming for a fully data-driven method, we use the following approach to define the respiratory gates:
\begin{enumerate}
    \item Split the acquired listmode data into short 0.5\,s time frames.
    \item Perform a low-resolution listmode OSEM without 
          modeling scatter nor attenuation for each short time frame.
    \item Use \replaced{principal}{princinple} component analysis to the 
          times series of the 
          3D reconstructions to  extract the respiratory gating signal, 
          similar to the approach proposed in \cite{thielemans2011device} 
          operating on a time series of sinograms. 
\end{enumerate}
Finally, amplitude-based gating was used to define 6 respiratory gates based on the extracted respiratory signal.

\subsection{PET data acquisition}

The performance of the proposed JMRA algorithms was evaluated by reconstructing 
a TOF PET acquisition of the anthropomorphic Wilhelm thorax phantom,  
as well as three patient acquisitions. 

\subsubsection{PET/CT acquisition of the anthropomorphic Wilhelm phantom}

%Listmode PET data of the dynamic human thorax phantom
%\cite{bolwin2018anthropomorphic} were acquired on a  Siemens Biograph mCT 
%TOF PET/CT system \cite{rausch2015} at Universit\"at M\"unster, Germany.
%The Wilhelm phantom simulates respiratory motion within a human-like thorax.
%Various compartments, including the liver, left ventricular myocardium, and
%torso, were filled with different [\textsuperscript{18}F]FDG 
%activity concentrations to mimic a realistic activity distribution. 
%Additionally, a small lesion was positioned near the diaphragm, 
%severely suffering from respiratory motion (motion amplitude 2~cm). 
%The phantom was prepared with a background activity concentration of 12.5~kBq/ml (thorax), 
%whereas the following activity concentration ratios were set accordingly: 
%myocardium/thorax: 6:1; lesion/thorax 20:1; liver/thorax: 2:1.
%In addition to a 30\,min listmode acquisition using realistic respiratory but no 
%cardiac motion, a second 10\,min scan without respiratory motion was acquired
%as reference. Both PET scans were performed about 8\,h after activity preparation.

Listmode PET data of the dynamic human thorax phantom 
\cite{bolwin2018anthropomorphic} were acquired on a Siemens Biograph mCT 
TOF PET/CT system \cite{rausch2015} at Universit\"at M\"unster, Germany. 
The Wilhelm phantom simulates respiratory motion within a human-like thorax. 
Various compartments, including the liver, left ventricular myocardium, and 
torso, were filled with different [\textsuperscript{18}F]FDG activity 
concentrations to mimic a realistic activity distribution. Additionally, a small 
lesion was positioned near the diaphragm, exhibiting severe respiratory motion 
(motion amplitude: 2~cm). 
The phantom was prepared with a background activity concentration of 
12.5~kBq/ml (thorax), and the following activity concentration ratios were set: 
myocardium/thorax: 6:1; lesion/thorax: 20:1; liver/thorax: 2:1. 
In addition to a 30\,min listmode acquisition with realistic 
respiratory motion \added{with a motion amplitude of 20\,mm in the
feet head direction at the liver dome}, 
a second 10\,min scan without respiratory motion was acquired as a reference. 
Both PET scans were conducted approximately 8\,h after activity preparation.

\subsubsection{Patient PET/CT acquisitions}

A low-dose attenuation CT scan during breath-hold, along with list-mode TOF PET data acquired 
in free-breathing mode, were collected from three patients using a 4-ring GE DMI PET/CT system \cite{hsu2017}.
In all cases a standard clinical whole-body acquisition protocol using an 
injected activity of 4.25\.MBq/kg [\textsuperscript{18}F]FDG and an acquisition
time of 80\,s per bed position 1\,h p.i. was used.
Patient details are summarized in Tab.~\ref{tab:patient_details}.

\added{All research was conducted in accordance with the principles embodied in the Declaration of Helsinki and in accordance with local statutory requirements.
All participants gave written informed consent to participate in this study.}

\begin{table}[t]
\renewcommand{\arraystretch}{1.3}
\small
\centering
\caption{Patient details for the data acquisitions using the 4-ring GE DMI PET/CT. 
In all cases \textsuperscript{18}F[FDG] and an acquisition time of 80\,s per bed position
was used.}
\label{tab:patient_details}
\begin{tabular}{ccccl}
\hline
\textbf{Patient} & \textbf{weight} & \textbf{inj. activity} & \textbf{acq. start p.i.} & \textbf{lesion location} \\ \hline
1                & 86\,kg                   & 366\,MBq      & 63\,min & subpleural lesion, right lower lobe \\
2                & 74\,kg                   & 322\,MBq      & 71\,min & subpleural lesion, left lower lobe \\
3                & 64\,kg                   & 272\,MBq      & 70\,min & lesion in lateral part of liver segment VIII \\ \hline
\end{tabular}
\end{table}

\subsection{Image reconstruction}

The data-driven estimated respiratory signal was used to define six 
respiratory gates by 
grouping the data according to the amplitude of the data-driven respiratory 
signal. To compare and benchmark the two proposed JRMA methods, the following 
reconstructions were performed for all datasets:
\begin{enumerate}
    \item OS-MLEM of the ungated emission data using static CT-based attenuation 
    correction without motion modeling. This reconstruction, called ``\textbf{no moco}", is expected 
    to suffer from motion and attenuation artifacts.
    \item JR MLEM using static CT-based attenuation correction and motion 
    warping operators derived from aligning gate-by-gate OS-MLEM reconstructions. 
    This reconstruction, called \textbf{JR MLEM}, is expected to mitigate motion artifacts 
    but still to suffer from attenuation mismatches.
    \item JR MLEM using gate-specific attenuation sinograms and motion warping 
    operators estimated from gate-by-gate MLACF reconstructions (called \textbf{Hybrid JRMA}).
    \item JR MLEM using gate-specific attenuation sinograms and motion warping 
    operators estimated through ADMM-based JRMA (called \textbf{ADMM JRMA}).
\end{enumerate}
To ensure a fair comparison across all methods, we used JR MLEM for the 
final image reconstruction in each case. Specifically, even when attenuation 
sinograms and motion warping operators were estimated via ADMM-based JRMA, 
the final single activity image was reconstructed using an  \replaced{additional}{addtional} JR OS-MLEM.
This approach allows for a direct comparison of all methods, 
independent of the image regularization used in ADMM-based JRMA, 
and provides a meaningful benchmark against standard clinical reconstructions.
For the Wilhelm phantom data set, we also performed an additional static OS-MLEM reconstruction
of the acquisition without respiratory motion and phase matched CT-based attenuation
correction that can serve as ground truth. 

\begin{figure*}
	\centering
	\includegraphics[width=1.0\textwidth]{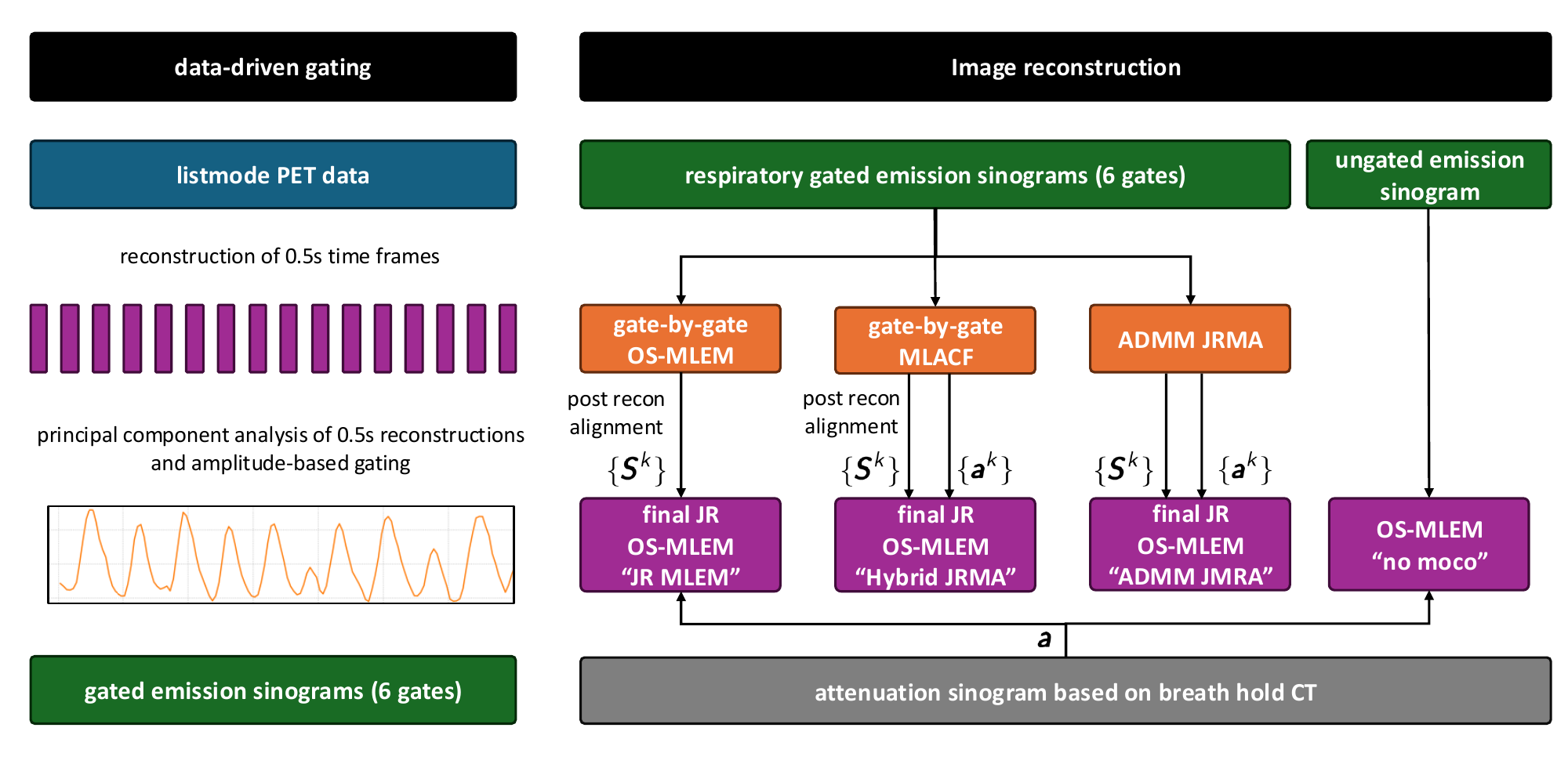}
	\caption{Overview of the workflow for data-driven gating and different image reconstruction
    algorithms used in this work. See text for details}
	\label{fig:workflow}
\end{figure*}

An overview of the processing workflow and reconstruction parameters is 
provided in Fig.~\ref{fig:workflow} and Tab.~\ref{tab:recon_params}.

\begin{table}[t]
\renewcommand{\arraystretch}{1.3}
\small
\centering
\caption{Overview of all reconstruction parameters}
\label{tab:recon_params}
\begin{tabularx}{\textwidth}{lX}
\hline
\textbf{reconstruction}  & \textbf{parameters} \\ \hline
gate-by-gate MLEM  & voxel size 3.6x3.6x2.8\,mm\textsuperscript{3}, 3 iterations 16 subsets \\ 
gate-by-gate MLACF & voxel size 5x5x5\,mm\textsuperscript{3}, 10 iterations 16 subsets, 3 attenuation updates
per activity update, $\gamma$ = 0.2 mean(emission sinogram)  \\ 
final JR MLEM      & voxel size 3.6x3.6x2.8\,mm\textsuperscript{3}, 3 iterations 16 subsets, 6\,mm FWHM Gaussian 
post filter \\
ADMM JRMA & voxel size 5x5x5\,mm\textsuperscript{3}, 50 outer iterations, $\rho$ = 10\textsuperscript{-6}\\
ADMM subproblem 1 & 3 iterations 16 subsets  \\
ADMM subproblem 2 & 1 update per subset activity update, $\gamma$ = 0.2 mean(emission sinogram) \\
ADMM subproblem 3 & 100 iterations, $\beta$ = 7$\cdot$10\textsuperscript{-5} \\ \hline
\end{tabularx}
\end{table}

\subsection{Image evaluation}

The evaluation of reconstructed images involved a side-by-side visual comparison 
to subjectively assess the reduction of motion and attenuation artifacts in coronal and sagittal slices as well
as in a coronal maximum intensity projection.
Particular attention was given to the presence of the ``banana attenuation artifact'' 
in the liver dome and to motion blurring in lesions affected by respiratory motion.

In addition to the visual assessment, the maximum activity concentration and contrast in lesions that 
were most affected by respiratory motion was compared between the different reconstructions.
The location of those lesions is given in Tab.~\ref{tab:patient_details}.
To calculate the contrast, the maximum activity concentration was divided by the local 
tissue background activity concentration.

\section{Results}
\label{sec:res}
\begin{figure}
    \centering
	\begin{subfigure}[b]{0.9\textwidth}
	   \centering
      \includegraphics[width=\linewidth]{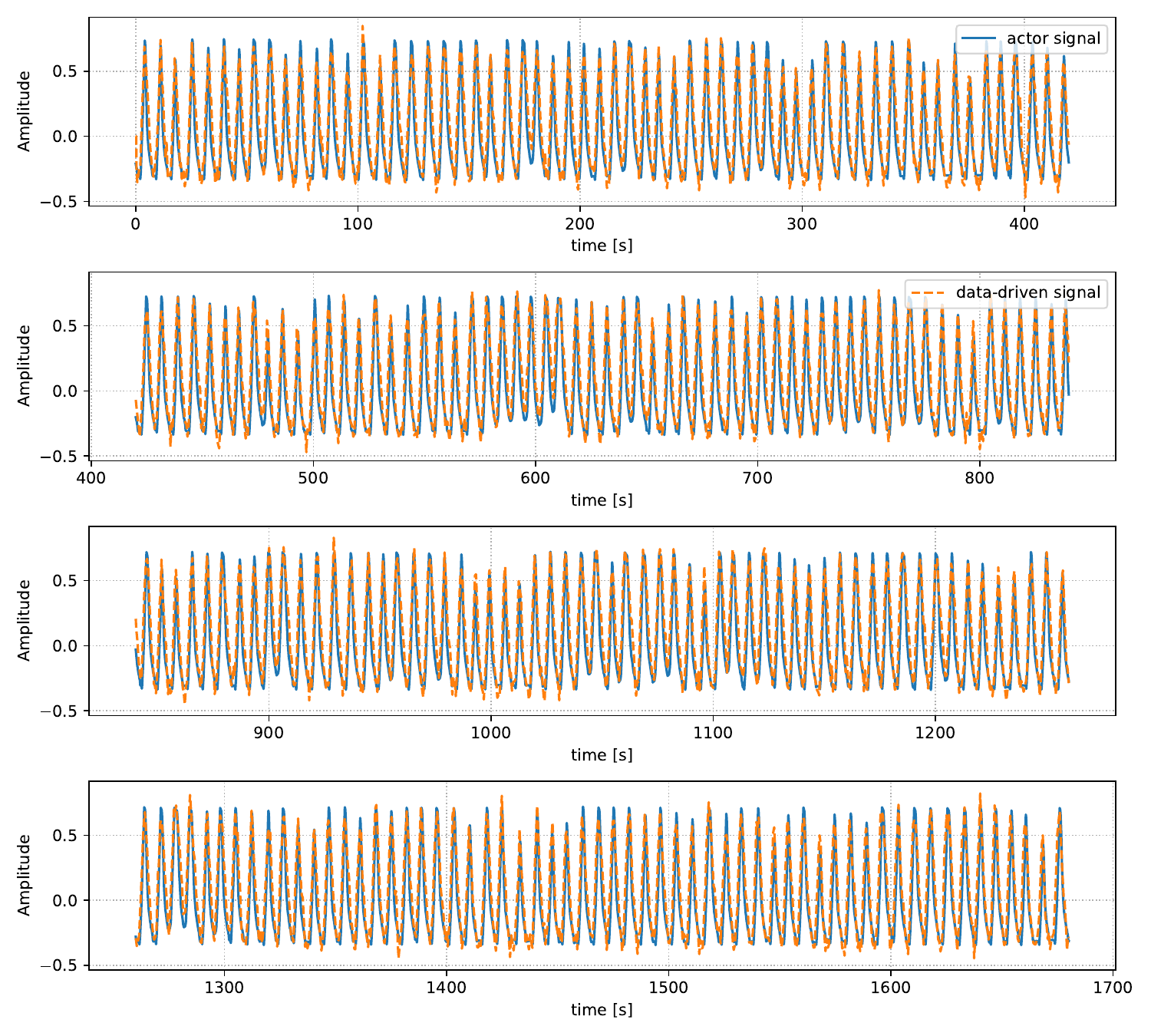}
      \caption{}
    \end{subfigure}
	\begin{subfigure}[b]{0.9\textwidth}
	   \centering
      \includegraphics[width=\linewidth]{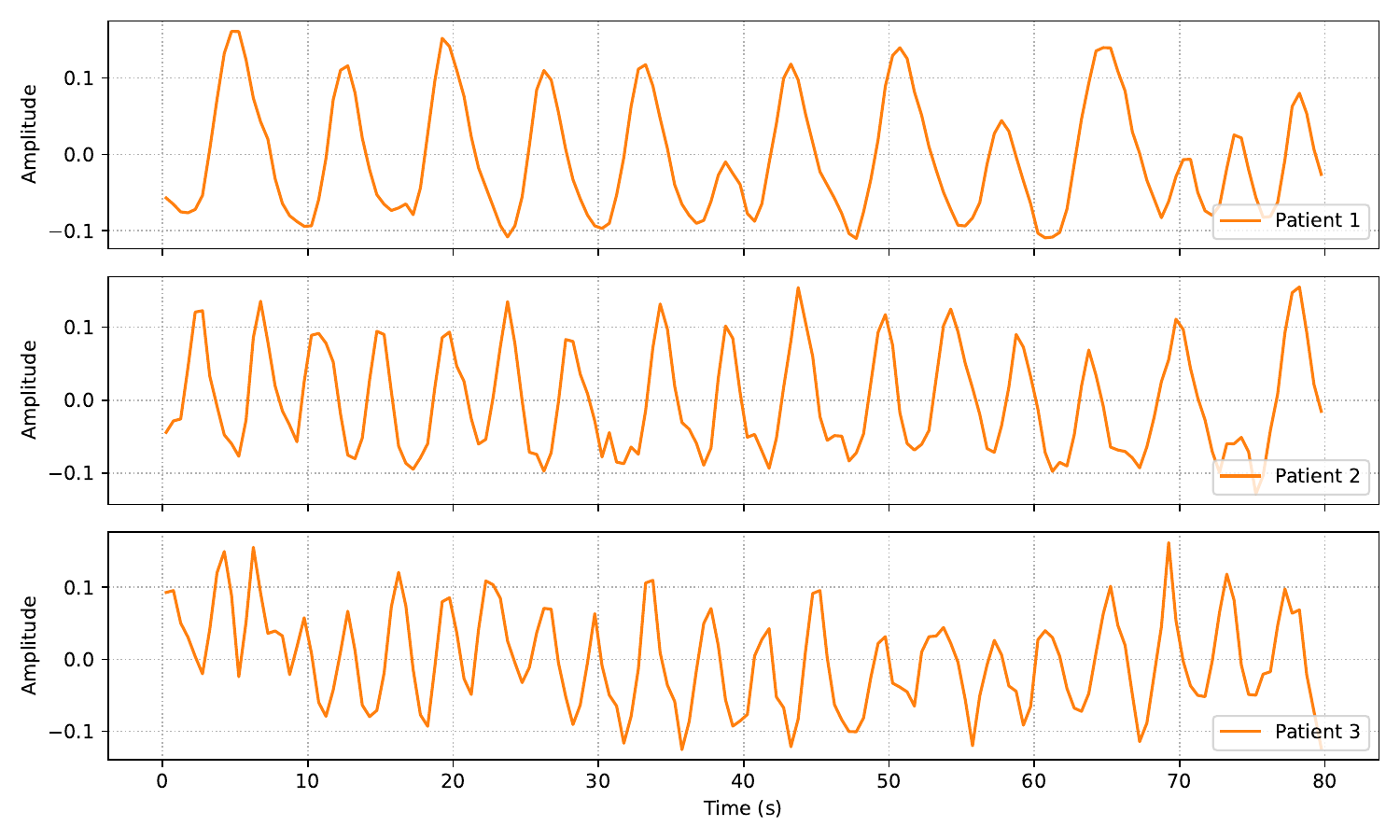}
      \caption{}
    \end{subfigure}
	\caption{PCA derived respiratory gating signal (orange) in comparison to actor 
    signal (blue) for (a) the Wilhelm and (b) the patient acquisitions.
    \added{In (a), the Pearson correlation coefficient between both signals is 0.91.}}
    \label{fig:resp_signal}
\end{figure}

Figure~\ref{fig:resp_signal} shows the PCA-based data-driven respiratory signal for
the Wilhelm phantom \replaced{and}{as well as for one of} the three patient acquisitions.
For the Wilhelm acquisition, shown in Fig.~\ref{fig:resp_signal}(a),
the data-driven respiratory signal was compared with the true motion signal
available from a sensor on the top of the actuator driving the respiratory motion.
The comparison demonstrates a close correlation 
\added{(Pearson correlation coefficient 0.91)} between the shapes of the signals 
indicating that the data-driven signal is well suited for respiratory gating.
Note that both signals are shown on the same arbitrary scale.
Since we observed almost perfect correlation between the two gating signals,
we decided to use actor-based ``ground-truth'' signal for the gate definition
in the Wilhelm acquisition.

For the patient case\added{s}, shown in Fig.~\ref{fig:resp_signal}(b),
a ground truth respiratory signal was not available.
However, a visual comparison with patient motion visible in the time series of
the short reconstructions
showed a close agreement between the respiratory motion
in the images and the data-driven respiratory signal.
For instance, a shallow breathing pattern was observed around 40 seconds and after 70
seconds in the images \added{of patient 1}, as well as in the data-driven signal.

\begin{figure*}
	\centering
	  \includegraphics[width=1.0\textwidth]{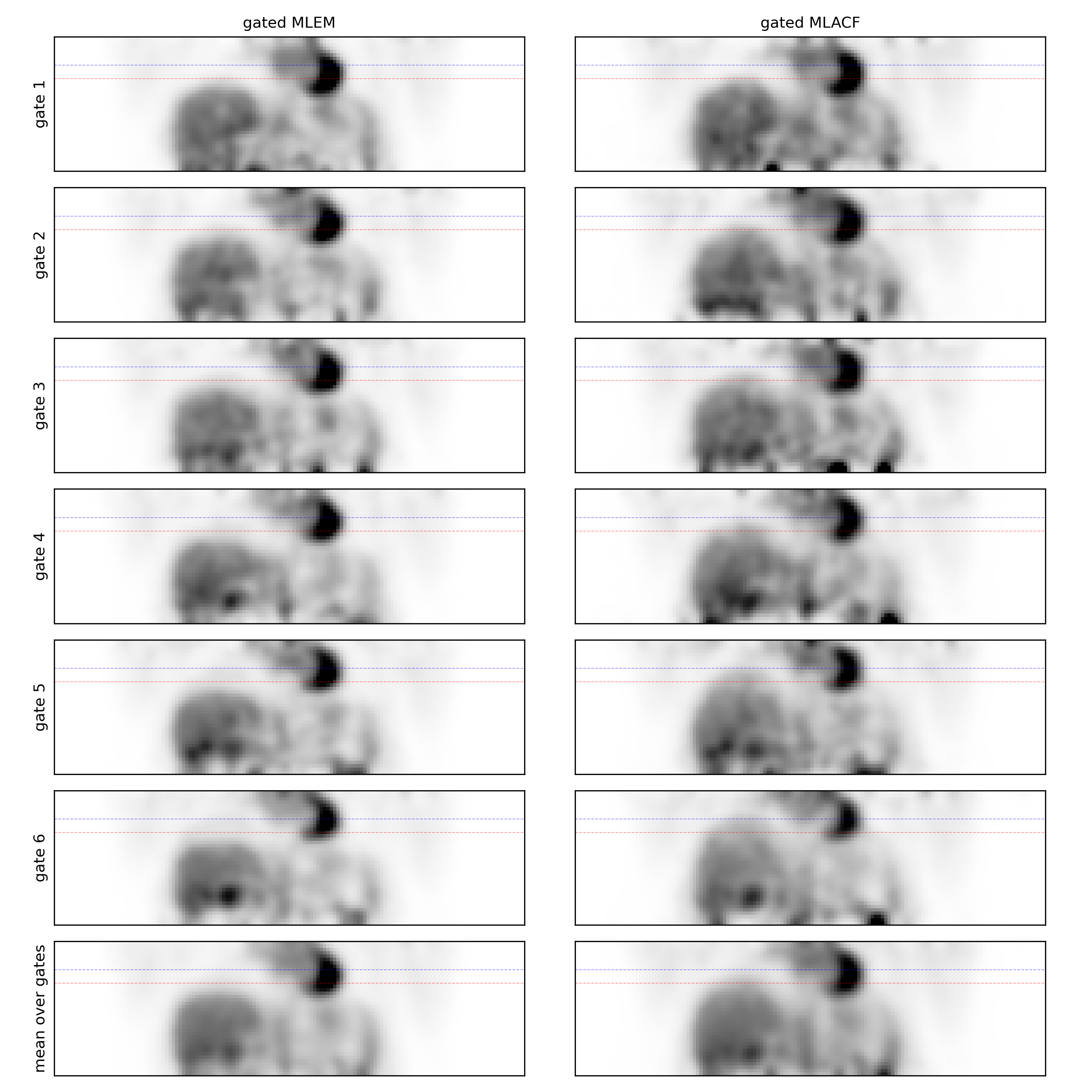}
	\caption{Same coronal slices of gate-by-gate MLEM reconstructions using a single static attenuation sinogram (left)
		and gate-by-gate MLACF reconstructions (right) reconstructions of the first patient data set.
		The horizontal dashed lines in red and blue indicate the position of
		the liver dome in the first and last respiratory gate.
		\deleted{The gated MLEM reconstructions clearly suffer from attenuation artifacts in the liver dome in almost all gates.}
        \added{The single static attenuation sinogram used in the MLEM 
        reconstructions leads to an attenuation artifact in the liver dome which is visible in almost all gates.
        Due to this attenuation artifact, it seems that the boundary between liver
        and right lung is almost unchanged between the gates.
        In contrast, the MLACF reconstructions using respiratory-matched 
        attenuation sinograms clearly show that, as expected, the liver lung 
        boundary is changing across gates.}
		\deleted{These artifacts are not present in the MLACF reconstructions.}}
	\label{fig:gated_recons}
\end{figure*}

%%%%
Figure~\ref{fig:gated_recons} shows the same post-smoothed coronal slice of 
gate-by-gate MLEM and MLACF reconstruction of the first patient data set.
The single static attenuation sinogram used in the MLEM reconstructions leads
to a clear attenuation artifact\deleted{s} in the liver dome which is visible in almost
all gates.
Due to this attenuation artifact, it seems that the boundary between liver
and right lung is almost unchanged between the gates which affects 
non-rigid alignment of the gated MLEM images.
In contrast, the gated MLACF reconstructions do not suffer from the 
attenuation mismatch resulting in change of liver lung boundary indicated
by blue and red dashed lines showing the liver dome position in end expiratory
and end inspiratory phase, respectively.

\begin{figure*}
	\centering
	\includegraphics[width=1.0\textwidth]{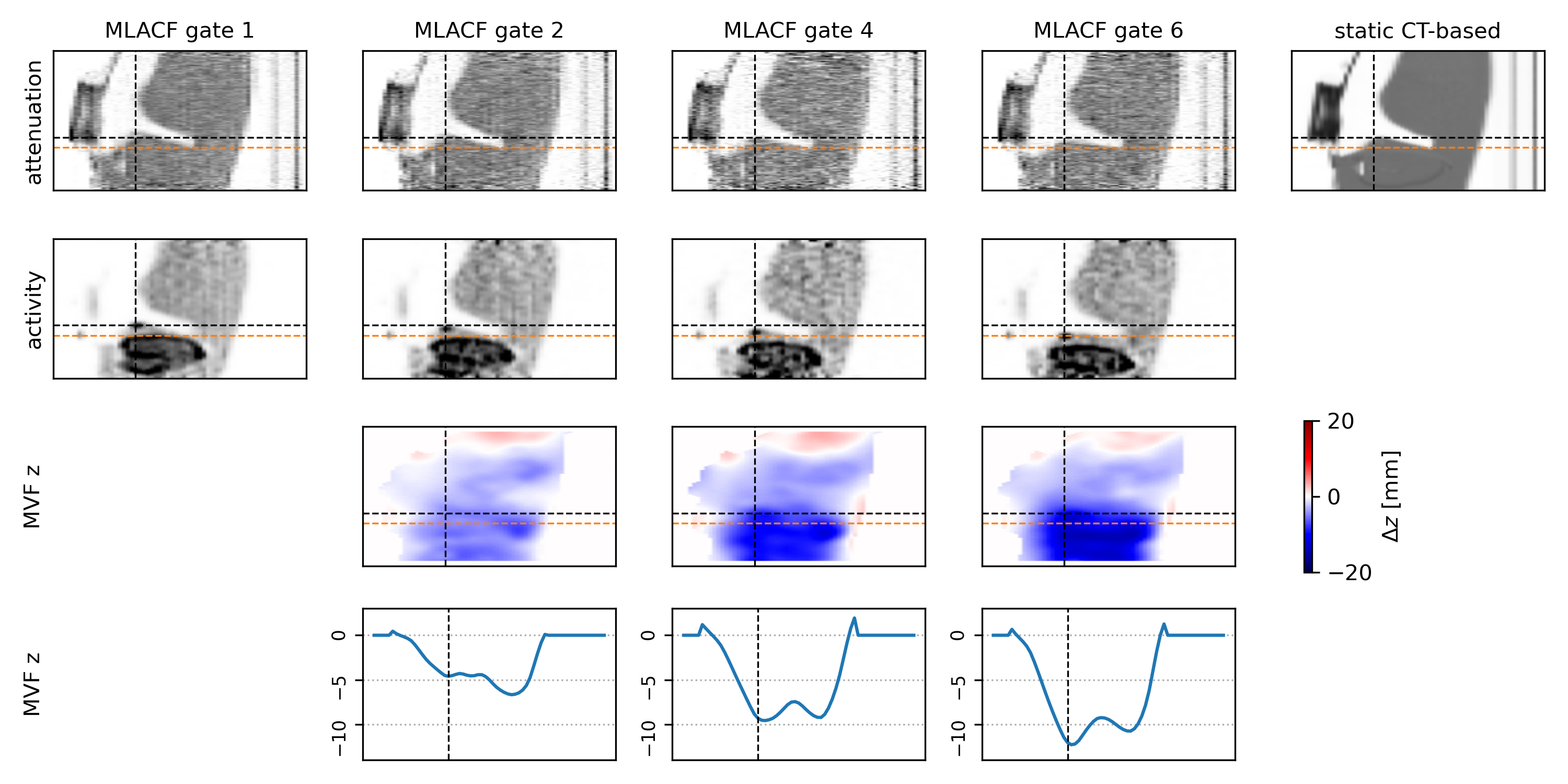}
	\caption{\added{(top row) sagittal slice of reconstructions of the estimated 
    gate-by-gate attenuation sinograms using MLACF in Hybrid JRMA and 
    static CT-based attenuation image used in 
    JR MLEM of the Wilhelm phantom acquisition shown for 4 out of 6 gates.
    Note that these images are reconstructions of the estimated attenuation sinograms and are shown for illustrative purposes only.
    (2nd row) corresponding gate-by-gate MLACF-based activity estimates.
    The horizontal black and orange lines indicate the position of
    the lesion at the liver dome in the first and last gate,
    respectively.
    Note the changing position of the liver dome across gates
    in MLACF-based attenuation images.
    (3rd row) z component of the estimated motion vector fields (MVFs)
    between the gates and the reference gate 1.
    (bottom row) profiles through the z component of the motion vector fields along the orange line.}}
	\label{fig:phantom_att}
\end{figure*}

%%%
\added{
Figure~\ref{fig:phantom_att} presents the activity distributions,
reconstructions of the estimated attenuation sinograms, and motion vector fields
from the gate-by-gate MLACF reconstructions used in Hybrid JRMA. 
The figure illustrates that the liver dome position in the reconstructed
MLACF estimated attenuation sinograms accurately reflects respiratory 
motion across the gates and aligns with the liver dome position in the 
corresponding activity images. 
Furthermore, it is evident that the static CT-based attenuation image,
acquired in the end-expiratory phase, does not align well with most of 
the gates. 
The estimated z-component of the motion vector field in the small lesion
at the liver dome between gate 6 and gate 1 is 12.2\,mm, based on the 
alignment of the gated MLACF activity reconstructions. 
In contrast, when aligning the gate-by-gate MLEM reconstructions, 
this value is reduced to 9.3\,mm.
}

%\added{
%Figure~\ref{fig:phantom_att} shows the activity distributions,  
%reconstructed estimated attenuation sinograms, and the estimated
%motion vector fields of the gate-by-gate MLACF
%reconstructions used in Hybrid JRMA.
%The figure demonstrates that the position of the liver dome in the
%reconstructed MLACF estimated  attenuation sinograms 
%reflects the respiratory motion across the gates and aligns with the
%position of the liver dome in the corresponding activity images.
%Moreover, it can be seen that the static CT-based attenuation image
%acquired in end expiratory phase used 
%does not align well with most of the gates.
%The estimated z component of the motion vector field between gate 6 and 
%gate 1 in the small lesion at the liver dome is 12.2\,mm based on the 
%alignment of the gated MLACF activity reconstructions.
%The same value was only 9.3\,mm when aligning the gate-by-gate MLEM reconstructions.
%}

\begin{figure*}
	\centering
	\includegraphics[width=1.0\textwidth]{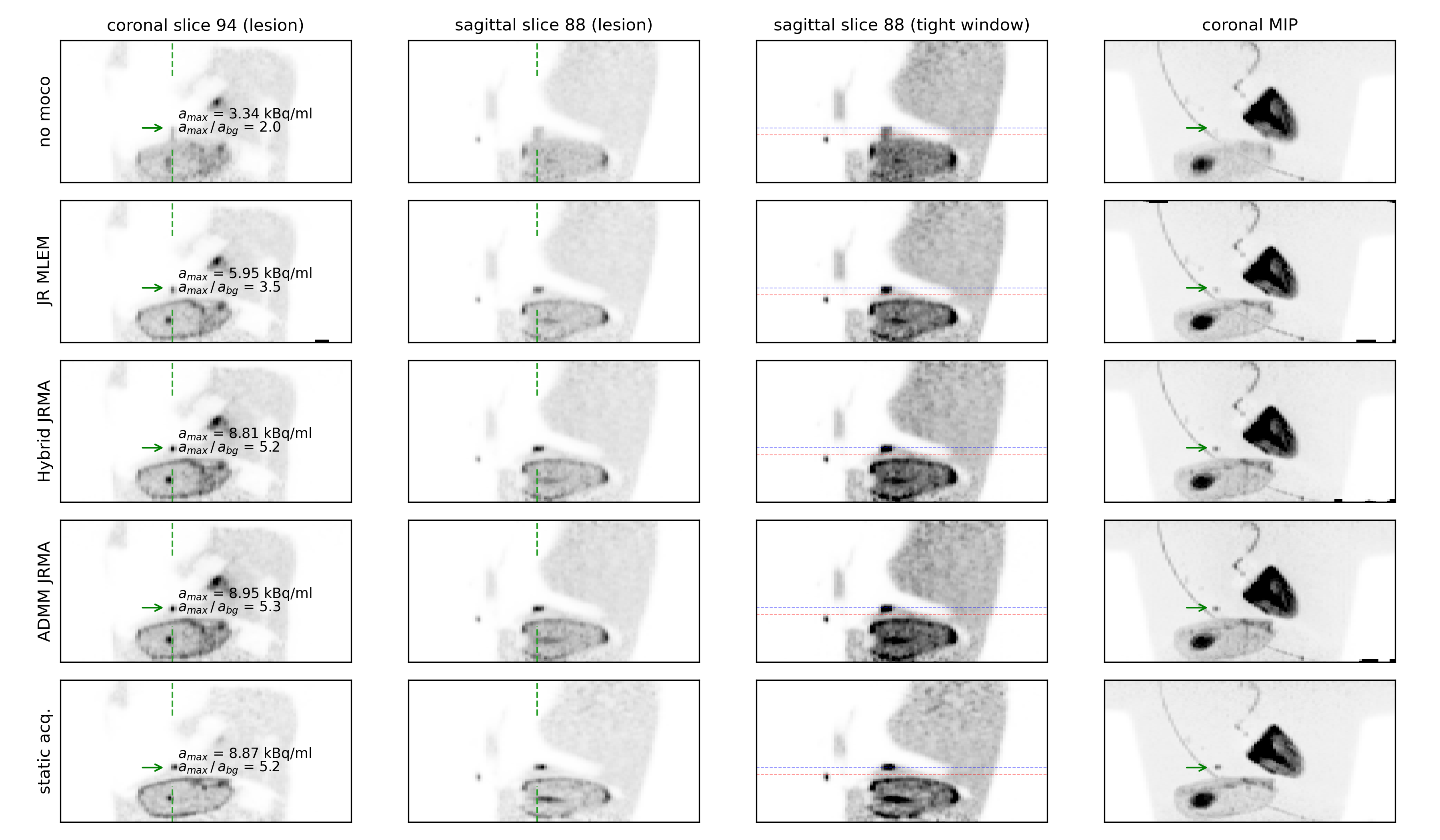}
	\caption{(left) coronal lesion slice, (2nd from left) sagittal lesion slice, (3rd from left) sagittal slice
		of liver dome in tighter color window, (right) coronal maximum intensity projection of:
		(top row) MLEM reconstructon of the moving anthropomorphic Wilhelm phantom without motion compensation,
		(2nd row) JR MLEM with single attenuation sinogram and motion warping operators from aligning gate-by-gate MLEM reconstructions
		(3rd row) JR MLEM with gate-by-gate attenuation and motion warping operators from aligning gate-by-gate MLACF reconstructions
		(4th row) JR MLEM with gate-by-gate attenuation and motion warping operators from ADMM-based JRMA
		(bottom row) MLEM reconstruction of static acquisition shown.
        The green dashed lines show the locations of the sagittal and coronal slices.
        The maximum activity concentration as well as the contrast of the lesion in the liver dome indicated by the green arrow 
        is also shown for comparison.
        }
	\label{fig:phantom}
\end{figure*}

Figure~\ref{fig:phantom} shows different reconstructions of the moving 
anthropomorphic Wilhelm phantom, as well as a MLEM reconstruction 
of the acquisition without motion.
JR MLEM is able to reduce the motion blur in the lesion in the liver dome
compared the the MLEM reconstruction without motion correction and increases
the lesion to background contrast from 2.0 to 3.5.
However, due to the attenuation mismatch, the contrast in this lesion is still 
underestimated compared  to the Hybrid JRMA and ADMM JRMA reconstructions which resulted in 
contrasts of 5.2 and 5.3, respectively.
The latter two values are in agreement with the lesion contrast of 5.2
observed in the MLEM reconstruction of the static acquisition.

\begin{figure*}
	\centering
	\includegraphics[width=1.0\textwidth]{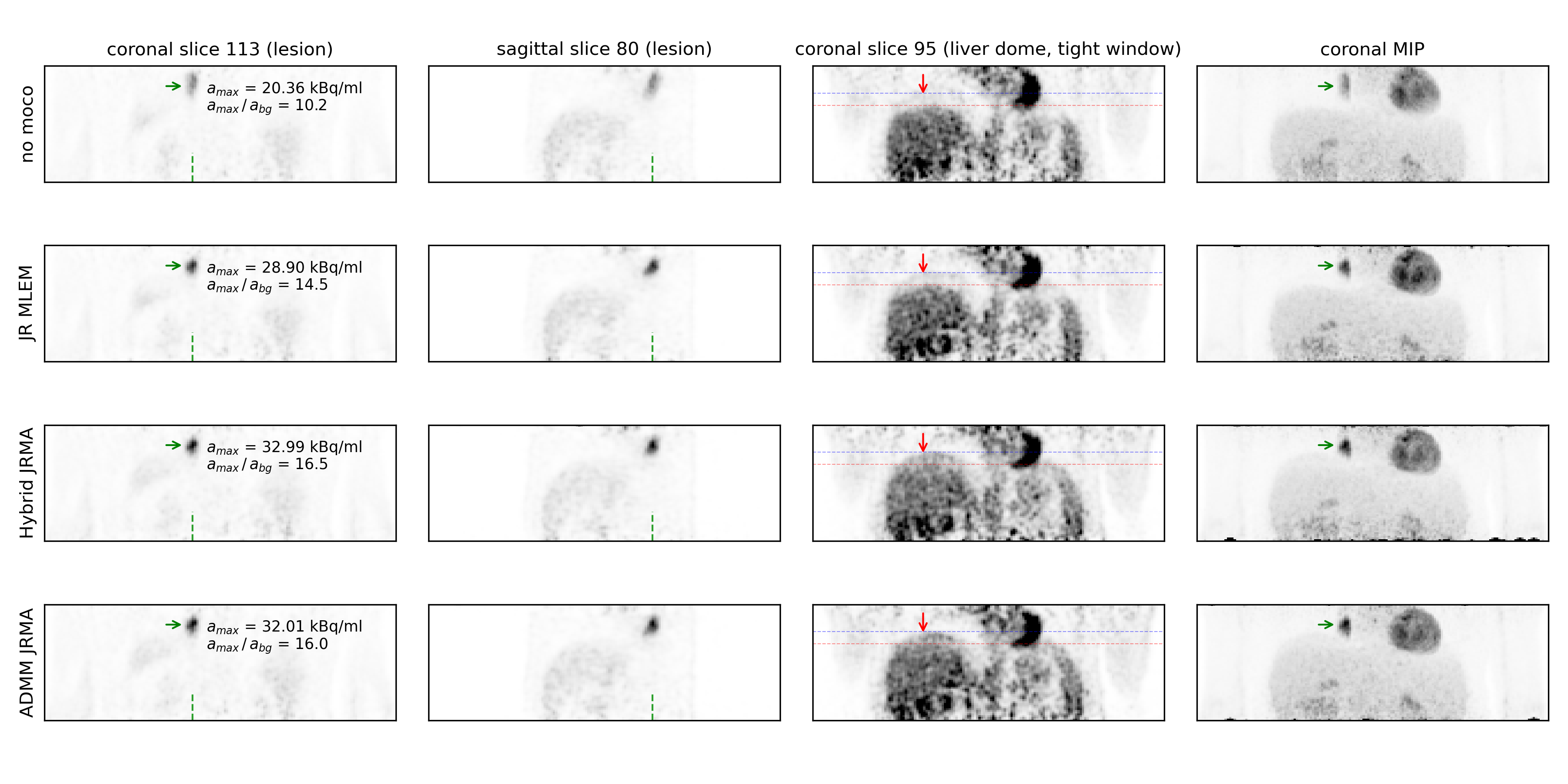}
	\caption{(left) coronal lesion slice, (2nd from left) sagittal lesion slice, (3rd from left) coronal slice of liver dome,
		(right) coronal maximum intensity projection of the first patient data set showing:
		(top row) MLEM reconstruction without motion compensation,
		(2nd row) JR MLEM with single attenuation sinogram and motion warping operators from aligning gate-by-gate MLEM reconstructions
		(3rd row) JR MLEM with gate-by-gate attenuation and motion warping operators from aligning gate-by-gate MLACF reconstructions
		(bottom row) JR MLEM with gate-by-gate attenuation and motion warping operators from ADMM-based JRMA.
        The green arrow indicated a lesion affected by respiratory motion.
        The left panel also shows the maximum activity concentration and the contrast of that lesion.
        The green dashed lines show the positions of the coronal and sagittal lesion slices.
        }
	\label{fig:p1}
\end{figure*}

\begin{figure*}
	\centering
	\includegraphics[width=1.0\textwidth]{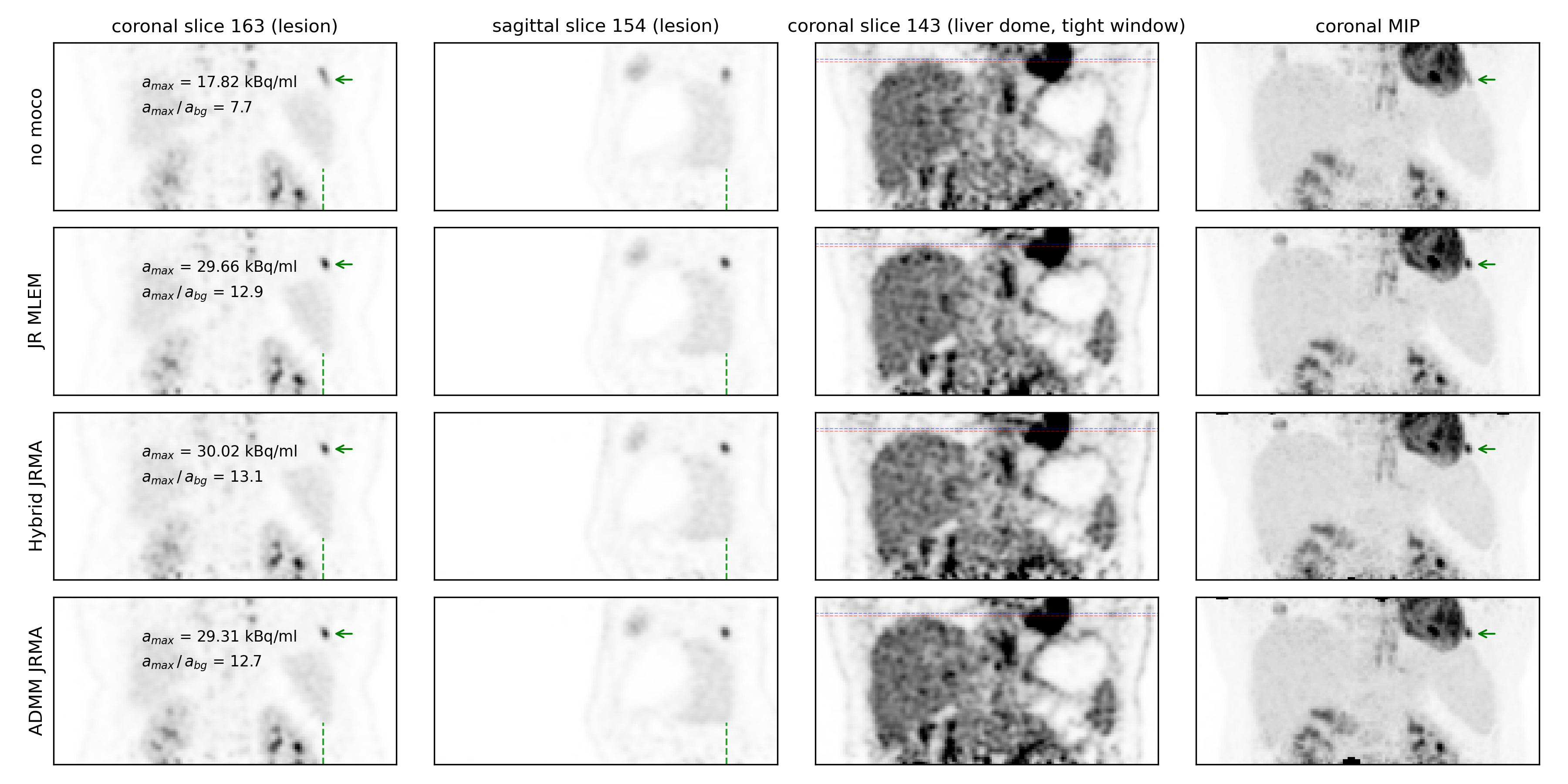}
	\caption{Same as Fig.~\ref{fig:p1} for the second patient data set.}
	\label{fig:p2}
\end{figure*}

\begin{figure*}
	\centering
	\includegraphics[width=1.0\textwidth]{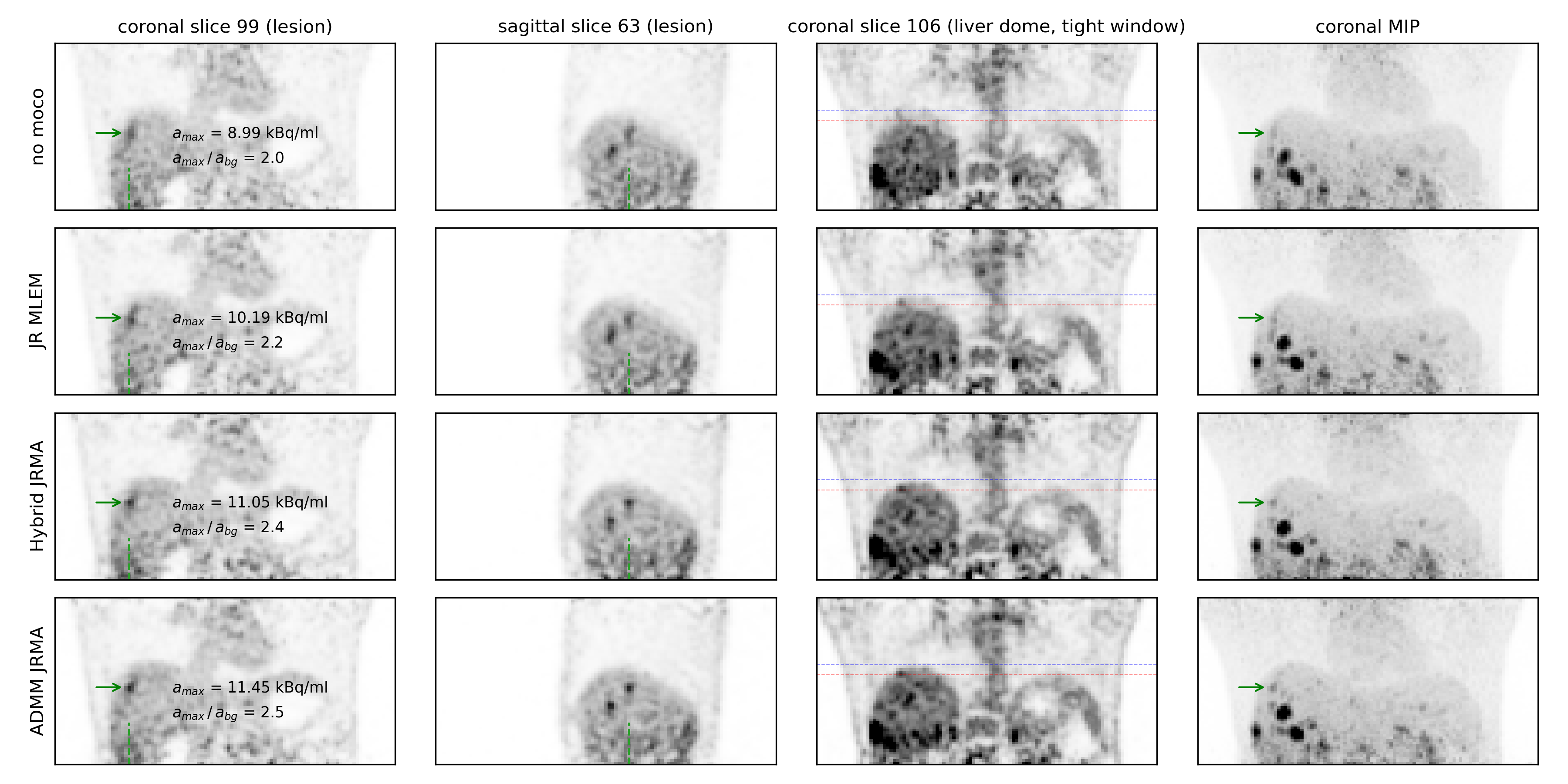}
	\caption{Same as Fig.~\ref{fig:p1} for the third patient data set.}
	\label{fig:p3}
\end{figure*}

%%%
% patient figs
Figures~\ref{fig:p1}, \ref{fig:p2}, and ~\ref{fig:p3} show the motion-uncorrected
and corrected images using three different techniques for the three patient data sets.
Respiratory motion results in a local gate-dependent activity and attenuation mismatch
when using a static CT-based attenuation sinogram in the MLEM reconstructions,
leading to the well-known banana artifact in the liver dome
as indicated by the red arrow in the first two rows of Fig.~\ref{fig:p1}.

Hybrid JRMA and ADMM JRMA successfully reduce these artifacts, as indicated in the bottom two
rows on Fig.~\ref{fig:p1}.
This improvement is primarily attributed to data-driven attenuation estimation using
MLACF in both joint methods, which use a gate-dependent attenuation estimate.

The quantification of the lesion contrast in the liver lesion
in Fig.~\ref{fig:p1} also demonstrates that motion correction with static 
CT-based attenuation leads to an increase in the lesion uptake compared
to the reconstruction without motion correction.
JR MLEM increases the lesion contrast from 10.2 to 14.5.
However, hybrid JRMA and ADMM JRMA show a higher lesion contrast of 
16.5 and 16.0, respectively.

In Fig.~\ref{fig:p2}, JR MLEM, hybrid JRMA, and ADMM JRMA increase the lesion
contrast from 7.7 to 12.9, 13.1, and 12.7, respectively, and strongly reduce
the motion blurring visible in the ``no moco'' reconstruction. 
In contrast\deleted{,} to the results of  Fig.~\ref{fig:p1}, the attenuation mismatch
in JR MLEM does not lead to a contrast underestimation.
This is likely because (a) the patient was breathing shallowly, and (b) the lesion 
is located in the left lung, at a distance from the diaphragm, resulting in a less 
pronounced attenuation mismatch between the gates.

In the small and low contrast liver lesion of Figure~\ref{fig:p3}, 
JR MLEM, hybrid JRMA and ADMM JRMA increase the lesion contrast from
2.0 to 2.2, 2.4, and 2.5, respectively.
Hybrid JMRA and ADMM JMRA mitigate the motion blurring, 
\replaced{whereas}{where} in JR MLEM
the lesion still suffers from motion blurring in the sagittal slice.

%%%%%%%%%%%%%%%%%%%%%%%%%%%%%%%%%%%%%%%%%%
% resp. trace 

%%%%%%%%%%%%%%%%%%%%%%%%%%%%%%%%%%%%%%%%
% gated MLEM vs MLACF

%%%%%%%%%%%%%%%%%%%%%%%%%%%%%%%%%%%%%%%%
% Wilhelm

%%%%%%%%%%%%%%%%%%%%%%%%%%%%%%%%%%%%%%%%
% P1

%%%%%%%%%%%%%%%%%%%%%%%%%%%%%%%%%%%%%%%%
% P2

%%%%%%%%%%%%%%%%%%%%%%%%%%%%%%%%%%%%%%%%
% P3

\section{Discussion}
\label{sec:dis}
In this study, we proposed and evaluated two approaches for the joint reconstruction 
of respiratory self-gated TOF PET data: the hybrid JRMA and the ADMM-based JRMA 
algorithms. Both methods aim to reconstruct a single activity image while accounting 
for the variations in attenuation across different respiratory gates. This is achieved 
through the joint estimation of activity, attenuation, and motion vector fields.

Our results demonstrate that the differences in image quality between the hybrid 
JRMA and the ADMM-based JRMA approaches are marginal. This outcome was unexpected, 
as we initially hypothesized that the ADMM-based JRMA would yield superior motion 
and attenuation estimates due to its more rigorous optimization framework.

A significant limitation of the ADMM-based JRMA algorithm is its sensitivity to 
the internal ADMM $\rho$ parameter. Reconstructions with varying $\rho$ values 
revealed that improper tuning of this parameter can lead to suboptimal performance 
or even divergence of the algorithm. This sensitivity is not surprising given that 
the ADMM modification applied here addresses a problem that is inherently non-convex.
A potential mitigation of the problem would be to investigate strategies to 
automatically tune the Lagrangian penalty parameter $\rho$ \cite{wohlberg2017}.

Given the small differences in image quality and the sensitivity of the ADMM-based 
JRMA to $\rho$, we recommend the use of the hybrid JRMA approach for practical 
applications. The hybrid method is simpler to implement and more robust, making 
it more suitable for routine clinical deployment.

Correctly handling the variations in attenuation due to respiratory motion is crucial 
in any joint reconstruction that uses all the acquired emission data. Failure to 
account for these changes can result in severe attenuation artifacts, such as the 
well-known banana artifacts observed in the liver dome. These artifacts not only 
degrade image quality but also adversely affect the estimation of motion and attenuation, 
potentially leading to inaccurate reconstructions. Our study underscores the importance 
of accurate attenuation correction as a critical component of joint reconstruction 
methods.

\added{
The quantitative analysis of the z-component of the motion vector fields estimated in
Hybrid JRMA for the Wilhelm acquisition indicates that the displacement of the lesion in
the liver dome is underestimated, measuring 12.2\,mm instead of the true 20\,mm. 
This underestimation can be attributed to three factors:
(i) Since the data was divided into six gates, the displacement of the lesion
between the first and last gate is reduced due to intra-gate motion.
(ii) The gate-by-gate MLACF reconstructions were performed using relatively thick slices (5\,mm) to reduce reconstruction time.
(iii) Regularization applied to the motion vector fields further smoothed 
the estimated displacement.
Notably, the underestimation was even more pronounced when aligning the gate-by-gate MLEM reconstructions with a single static CT-based attenuation image, where 
the estimated displacement was only 9.3\,mm instead of the true 20\,mm.
}

The hybrid JRMA algorithm potentially holds high clinical relevance. 
It enables the reconstruction of a single activity image from all acquired counts, 
thus minimizing noise while mitigating motion artifacts.
By  accurately correcting for photon attenuation, our method produces high-quality 
diagnostic images, making it a practical and impactful tool that should be
further investigated for routine clinical use.

A notable strength of our work is the evaluation of the proposed methods on data 
from two different state-of-the-art TOF PET scanners. This demonstrates the 
robustness and generalizability of our algorithms across different scanner 
architectures, a feature rarely explored in similar studies.

\subsection{Future Research Directions}

Future research should focus on improving the non-rigid motion estimation between 
gated PET images. Currently, we use a basic demons algorithm that relies solely 
on local smoothness as prior information. Incorporating more sophisticated priors, 
such as learned priors for respiratory motion fields, could enhance the accuracy 
of motion estimation.

Another area for improvement is the tuning of the intensity prior applied 
to the correction factors in the MLACF-based attenuation estimation. 
On the one hand, it must be strong enough to resolve the scale problem 
by encouraging correction factors for pure tissue lines-of-response 
to approach unity. 
On the other hand, it should not be too strong, 
as this could suppress local differences in the attenuation sinogram 
caused by respiratory motion.

\subsection{Limitations}

There are several limitations to our study. First, due to computational constraints, 
we processed data from only a single bed position in multi-bed acquisitions. 
However, extending the proposed methods to multi-bed acquisitions should pose 
no fundamental issues.
Second, our proof-of-concept implementation is computationally intensive, with 
long processing times. Future work should aim to accelerate these reconstructions 
by leveraging more efficient projectors and registration algorithms, potentially 
running on GPUs.
Another limitation of this study is the use of a static scatter estimate rather 
than gated scatter estimation. However, since scatter is spatially smooth\deleted{ relative}, 
we do not anticipate this to cause significant issues in the reconstructed images.
Finally, our current implementation operates on sinogram data, which is suboptimal 
for gated datasets with low count statistics. Extending MLACF to handle list-mode 
data could address this issue, although this would introduce additional complexities.

\section{Conclusion}
\label{sec:con}
This paper presents two novel algorithms, Hybrid JRMA and ADMM-based JRMA, for the joint estimation of activity, attenuation, and respiratory motion in TOF PET imaging. These methods enable the reconstruction of a single activity image that is both motion-corrected and free from attenuation artifacts, without the need for external hardware. Given its greater robustness, simplicity, and comparable performance to the more complex ADMM-based JRMA, the Hybrid JRMA algorithm is particularly well-suited for clinical implementation.

\section*{Acknowledgments}
\label{sec:ack}
This work was supported by NIH grant R01EB029306 
\added{and FWO project G062220N}.

\appendix
\section{Solutions to the four subproblems of the ADMM-based JRMA}
\label{app:one}
\subsection{Solution of JRMA ADMM subproblem 1}

Subproblem 1 can be solved gate-by-gate and includes optimization of the negative Poisson log likelihood and a quadratic penalty
\begin{equation}
	\bm{z}^{k,(n+1)} = \argmin_{\bm{z}^k} \mathcal{D}_k(\bm{z}^k, \bm{a}^{k,(n)}) + \frac{\rho}{2} \Vert \bm{z}^k -
	\underbrace{(\bm{S}^{k,(n)} \bm{\lambda^{(n)}} - \bm{u}^{k,(n)})}_{\bm{b}^{k,(n)}} \Vert_2^2 \label{eq:sp1} \ ,
\end{equation}
where \eqref{eq:sp1} can be solved efficiently using the iterative algorithm proposed by De Pierro~\cite{de1995modified}, using the update
\begin{equation}
	\bm{z}^{k}_{(m+1)} = \frac{2\bm{v}}{\sqrt{\bm{w}^2+4\rho \bm{v}} + \bm{w}} \ ,
\end{equation}
with
\begin{equation}
	v = \bm{z}^{k}_{(m)} \bm{P}^T {\bm{a}^{k,(n)}} \frac{\bm{y}^k}{\bar{\bm{y}}(\bm{z}^k_{(m)})} \text{~and~} w = \bm{P}^T
	{\bm{a}^{k,(n)}} - \rho \, \bm{b}^{k,(n)} \ ,
\end{equation}
where the subscript $(m)$ denotes the ``inner'' iteration of the De Pierro algorithm.
We initialize $\bm{z}^{k}_{(0)}$ with $\bm{z}^{k,(n)}$.

\subsection{Solution of JRMA ADMM subproblem 2} \label{sec:app_mlacf}

An approach to estimate the attenuation sinograms for every gate in the presence of a CT scan
performed in breath hold, is to write the attenuation sinograms for every gate as a product of the
breath hold (static) CT-based attenuation sinogram $\tilde{\bm{a}}$ and gate dependent ``correction factors"
$\bm{g^k}$ that account for the respiratory motion mismatch between the gates and the CT acquisition
\begin{equation}
	\bm{a^k} =  \bm{g^k} \tilde{\bm{a}} \ .
\end{equation}
This has the advantage that for most LORs - especially for the LORs that are not or little affected
by respiratory motion - these correction factors should be close to 1.
Consequently, the gate-by-gate attenuation sinogram correction factors can be updated based on
current reconstruction $\bm{\lambda}^{(n)}$, the current motion deformation operators $\bm{S^{k,(n)}}$,
and the acquired gated emission data $\bm{y_k}$ by solving the optimization problem
\begin{equation}
	\bm{g^{k,(n+1)}} = \argmin_{\bm{g}^k} \mathcal{D}_k(\bm{S^{k,(n)}}\bm{\lambda}^{(n)}, \bm{g}^k \, \tilde{\bm{a}}) +
	\underbrace{\gamma \Vert 1 - \bm{g}^k \Vert_2^2}_{\mathcal{R}_2(\bm{g}^k)} + \chi_{\geq 0}(\bm{g}^k) \ , \label{eq:mlacf_problem}
\end{equation}
favoring values of $g_k$ that are close to unity due to intensity prior $\mathcal{R}_2(\bm{g})$.
If $\mathcal{D}_k$ is the Poisson log likelihood and the additive contaminations (the expectation of scattered and random coincidences)
in the forward model are non zero, \eqref{eq:mlacf_problem} has no analytic solution.

To obtain an analytic solution, we approximate $\mathcal{D}_k$ with a non-TOF weighted Gaussian log likelihood,
inspired by \cite{rezaei2014mlacf} leading to the analytic MLACF update
\begin{align}\label{eq:attclosedform}
	g_{i}^{k,(n+1)}  & = \left(\dfrac{\sum_t (y_{it}^k - r_{it}^k) + \gamma \sum_t y_{it}^k / \sum_t q^{k,(n)}_{it}}
	{\sum_t q^{k,(n)}_{it} + \gamma \sum_t y_{it}^k / \sum_t q^{k,(n)}_{it}} \right)_+                               \\
	\bm{a}^{k,(n+1)} & = \bm{g}^{k,(n+1)} \tilde{\bm{a}} \ ,
\end{align}
where $\bm{q}^{k,(n)} = \tilde{\bm{a}} \bm{P} \bm{S}^{k,(n)} \bm{\lambda}^{(n)}$.

\subsection{Solution of JRMA ADMM subproblem 3}

Subproblem \eqref{eq:sp_3} is related to a denoising problem that tries to combine the gate-by-gate
images $\bm{z}^k + \bm{u}^k$ into a single activity image $\bm{\lambda}$ combined with a regularizer
on $\bm{\lambda}$
\begin{equation}
	\bm{\lambda}^{(n)+1} = \argmin_{\bm{\lambda}} \sum_k \frac{\rho}{2} \Vert \bm{S}^{k,(n)} \bm{\lambda} - (\bm{z}^{k,(n+1)} +
	\bm{u}^{k,(n)}) \Vert_2^2 + \mathcal{R}_1(\bm{\lambda}) \label{eq:denoise_orig} \ ,
\end{equation}
that in principle can be solved using gradient descent or proximal gradient descent depending on the
smoothness of the regularizer $\mathcal{R}_1$.
Note, however, that to evaluate the gradient of the first term, an evaluation of the all image warping
operators $\bm{S}^k$ and their adjoints ${\bm{S}^k}^T$ are needed in every update which can be become
very time consuming.

To obtain a computationally more efficient solution of \eqref{eq:denoise_orig}, we approximate problem
\eqref{eq:denoise_orig} by
\begin{equation}
	\bm{\lambda}^{(n+1)} = \argmin_{\bm{\lambda}} \frac{\rho}{2\sigma}\Vert \bm{\lambda} - \bm{v}^{(n)} \Vert_2^2 + \mathcal{R}_1(\bm{\lambda})  \label{eq:denoise} \ ,
\end{equation}
with
\begin{equation}
	\bm{v}^{(n)} = \frac{1}{n_k} \sum_{k=1}^{n_k} \left(\bm{S}^{k,(n)}\right)^{-1} \Big(\bm{z}^{k,(n+1)} + \bm{u}^{k,(n)} \Big)
\end{equation}
where ${\bm{S}^k}^{-1}$ represents the inverse of $\bm{S}_k$, and $n_k$ is the number of respiratory gates.
To achieve edge-preserving noise supression which is important for motion estimation, we use a total variation
\begin{equation}
	\mathcal{R}_1(\bm{\lambda}) =  \beta \Vert \bm{\nabla} \bm{\lambda} \Vert_{2,1}
\end{equation}
as regularizer acting on $\bm{\lambda}$.
The corresponding non-smooth optimization problem \eqref{eq:denoise} was solved iteratively using
the fast gradient projection method proposed by Beck and Teboulle~\cite{beck2009fast}.

\subsection{Solution of JRMA ADMM subproblem 4}

Subproblem \eqref{eq:sp_4} is a non-ridig registration task, aligning individual gated reconstructions $\bm{z}^k$ to
the $\bm{z}^{k_\text{ref}}$ of the reference gate to estimate the corresponding deformation warping operator $\bm{S}^k$ 
between each gate and the reference gate.
This subproblem can be solved gate-by-gate by minimizing the cost function
\begin{eqnarray}
	\bm{S}^{k,(n+1)} = \argmin_{\bm{S}^k} \Vert \bm{S}^k \bm{z}^{k,(n+1)} - \bm{z}^{k_\text{ref},(n+1)} \Vert_2^2 +
	\mathcal{R}_3(\bm{S}^k) \label{eq:align}
\end{eqnarray}
In this work, the solution of minimization subproblem \eqref{eq:align} is approximated using a in-house
implementation of a regularized version of the diffeomorphic Demon's algorithm
\cite{vercauteren2009diffeomorphic} resulting in updated displacement vector fields that are used in
motion warping operators $\{\bm{S}^k\}$.

\printbibliography

\end{document}